# Decoding AI Tutor Effects for Educational Measurement: Temporal, Multi-Outcome, and Behavior-Cognitive Analysis

Yiyao Yang
0009-0001-8693-4888
yy3555@tc.columbia.edu
Columbia University, New York, NY, United States

Yasemin Gulbahar
0000-0002-1726-3224
yg2918@tc.columbia.edu
Columbia University, New York, NY, United States

## I. Abstract

Artificial intelligence (AI) tutors have become increasingly popular in learning environments. In this study, we propose an AI agent prototype framework for exploring AI-assisted learning with temporal interaction patterns, multiple outcomes analysis, and behavioral-cognitive learner profiling. Based on three research questions, this study aims to investigate whether early interaction patterns can predict later performance and trust, how multiple outcomes can be traded off with different AI tutor feedback conditions, and if learner profiles can be identified with behavioral and cognitive indicators. An AI tutor agent has been developed to provide various feedback forms to learners, including hints, explanations, examples, and code. A neural policy model and a stochastic simulation framework are used to produce artificial student-AI tutor interaction records, which include response time, attempts, hint requests, correctness, quiz results, improvement, satisfaction, and trust. Temporal features are used to predict later correctness and trust with early interaction patterns, and clustering methods are used to find learner profiles. The results showed that early interaction patterns were predictive of later performance and trust, that student behavior changed over time with AI-based tutoring, and that latent student profiles could be identified based on their behavioral and cognitive differences.

## II. Introduction

Artificial intelligence (AI) tutors have become more prevalent as part of the digital learning environment, and their capabilities to offer guidance and support to learners have been continuously improved. With the recent breakthroughs of large language models and intelligent tutoring systems, AI tutors have expanded their capabilities to offer support for complex tasks, and their use is likely to become more prevalent. As the use of AI tutors becomes more prevalent, the measurement of the interaction between the learners and the AI tutor and the interpretation of the interaction have become a significant challenge for educational measurement and learning analytics research.

Unlike the use of traditional assessment tools, where the responses of the learners to the test questions remain the same, the use of AI-assisted learning environments provides a rich source of interaction data. These interaction data consist of various behavioral responses, and previous studies using educational data mining have shown that the process data contain significant features of the learning process, such as learners' engagement and persistence (Baker & Inventado, 2014). However, the use of AI tutors as a significant part of the learning environment presents a new measurement challenge for the learning process. Researchers must consider the interpretation of the learning process, as the interaction between the learners and the AI tutor provides a new source of process data.

One of the key challenges relates to the temporal learning dynamics in the context of AI assistance. It is often the case that learners engage in multiple turns of dialogue or feedback cycles in the context of an AI tutoring system, and their behavior can vary over time. The early interaction signals, such as response latency or help-seeking, can be predictive of later learning outcomes. The study of intelligent tutoring systems has shown that the temporal dynamics of learner interaction can be an early indicator of learner success or difficulty (Turvey, 1990). The question of whether early interaction signals can be predictive of later learner performance and trust in the AI tutoring system is an important research question in the context of educational measurement and learning systems.

The second challenge relates to the multi-outcome nature of learner performance in the context of an AI tutoring system. Unlike traditional educational measurement, where learner performance is often single-dimensional, learning in the context of an AI tutoring system does not yield single outcomes, and there are multiple dimensions of performance and perception, such as correctness, improvement over time, perceived usefulness, satisfaction, and trust in the AI tutoring system. The educational measurement framework needs to account for trade-offs between multiple outcomes. For instance, highly directive feedback can enhance correctness but compromise learner autonomy and trust in the system, and exploratory feedback can enhance deep learning but compromise the rate of performance improvement. The multi-dimensional outcome structure in the context of an AI tutoring system can be captured by multi-dimensional outcome modeling approaches.

The third challenge relates to the learner heterogeneity in the context of an AI tutoring system. Learners vary in their prior knowledge, programming experience, and interaction strategies in the context of an AI tutoring system. The differences in learner behavior and cognition can be reflected in distinct learner engagement and outcome patterns across learners. Identifying learner profiles based on their behavioral and cognitive indicators can provide valuable insights into how learners benefit from an AI tutoring system. Clustering and learner profiling approaches have been widely used in the context of learning analytics to uncover learner outcome and interaction patterns (D'Mello et al., 2017).

To tackle these challenges, this study offers a prototype research framework for analyzing AI tutor interactions from an educational measurement perspective. Specifically, the proposed framework includes an AI tutor agent, synthetic student-AI tutor interaction logs, and analytical models to capture the temporal dynamics, multi-outcome trade-offs, and behavioral-cognitive learner profile. By incorporating the ideas from educational measurement, learning analytics, and AI tutor interaction, this study seeks to develop a methodological underpinning for analyzing interaction data with an AI tutor and conducting future empirical research with real-world AI tutor interaction data.

III. Literature Review

**AI Tutors and Educational Measurement**

AI tutors and intelligent tutoring systems (ITS) have emerged as significant tools in the context of digital learning environments. Previous research on ITS has shown that AI-based tutors have the potential to improve the performance of students. For example, research has shown that AI-based tutors can improve the performance of students with the help of step-by-step guidance and feedback while engaging in problem-solving activities (VanLEHN, 2011; Chrysafiadi et al., 2022). Recent advancements in machine learning and language models have further enhanced the role of AI-based tutors. For example, recent research has shown that AI-based tutors have the potential to support complex tasks such as programming, data analysis, and problem-solving activities (Kasneci et al., 2023; Zawacki-Richter et al., 2019). In the context of educational measurement, the use of AI-based tutors has the potential to create significant opportunities. For example, the use of AI-based tutors has the potential to create significant amounts of interaction data between learners and intelligent tutors. Previous research on educational data mining and learning analytics has shown that process data has the potential to reveal significant patterns of learning behavior that may not be observed with the help of traditional assessment tools (Baker & Inventado, 2014; Romero & Ventura, 2010; Siemens & Baker, 2012). With the growing use of AI-based tutors, the question that arises is how the interaction data can be used to measure the performance of learners.

**Temporal Interaction Dynamics in AI-Assisted Learning**

One important area of research concerns the temporal dynamics of student–AI interactions. Unlike traditional testing environments, where responses are typically independent, AI-assisted learning tasks involve iterative interactions between students and the tutoring system. Students may request hints, revise their answers, and receive feedback across multiple interaction turns. As a result, learning behaviors unfold over time and may exhibit temporal patterns that provide insight into student learning processes.

Prior studies in intelligent tutoring systems have shown that temporal interaction features, such as response latency, help-seeking patterns, and sequences of student actions, can be used to predict learning outcomes and detect students who may be struggling (Koedinger et al., 2015; Aleven et al., 2003). Similarly, research in learning analytics has emphasized the importance of analyzing interaction traces to identify early indicators of learning success or difficulty (Ferguson, 2012; Roll & Winne, 2015). Temporal behavioral signals may therefore provide useful predictive information about later performance and student perceptions of AI tutors, including trust in automated feedback systems (D'Mello et al., 2017).

Despite these advances, relatively few studies have examined how early interaction behaviors in AI-assisted tasks relate to multiple downstream outcomes, particularly when students interact with conversational or generative AI tutors. Understanding the predictive role of early temporal interaction signals remains an important challenge for educational measurement in AI-supported learning environments.

**Multi-Outcome Measurement in AI-Assisted Learning**

Another major challenge relates to the multi-dimensional nature of learning outcomes within an AI-supported learning environment. Conventional approaches to learning measurement are based on a single outcome variable. However, learning with an AI tutor might result in a number of learning outcomes. These might include improvement in performance, usefulness of feedback provided by the AI tutor, satisfaction with learning, and trust in the learning process with an AI tutor.

Learning analytics research has focused on the significance of learning outcome modeling to capture the complexity of digital learning processes (Tempelaar et al., 2015; Henrie et al., 2015). Learning with an AI tutor might result in a number of learning outcomes. These might include improvement in performance, usefulness of feedback provided by the AI tutor, satisfaction with learning, and trust in the learning process with an AI tutor. Therefore, analyzing these learning outcomes will be critical to understanding the learning process with an AI tutor.

The recent research on learning with an AI tutor has focused on the importance of trust and reliability when a student uses an AI tutor. Students' trust in an AI tutor might influence the way a student uses the feedback provided by an AI tutor (Kasneci et al., 2023). Therefore, analyzing learning outcomes with perception-based learning results will be critical to understanding the learning process with an AI tutor.

**Behavioral-Cognitive Learner Profiling**

Apart from temporal interaction patterns and multi-outcome modeling, learner heterogeneity is another aspect that is being taken into account in the research on AI-Assisted Learning. Learners differ in their prior knowledge, motivation, and learning and interaction strategies in an AI-Assisted environment. These differences could lead to differences in the learning paths and outcomes of different learners.

Research in learning analytics has employed clustering and learner profiling techniques to detect different groups of learners based on their behavioral and cognitive attributes (Romero & Ventura, 2010; Tempelaar et al., 2015). These techniques could help in identifying learners' profiles, such as highly engaged learners, help-seeking learners, or disengaged learners, and so on, and how they could interact with an AI-Assisted environment. Identifying learners' profiles could offer an opportunity to understand how learners benefit from an AI-Assisted environment and how the feedback strategies of an AI-Assisted environment could be tailored to meet the needs of different learners.

In the case of an AI-Assisted environment, learners' behavioral-cognitive profiling could offer an opportunity to understand how learners' characteristics, such as their prior knowledge and motivation, and their behavioral attributes, such as their response times and hint requests, could influence their learning outcomes. This perspective is in accordance with recent research that emphasizes the importance of combining learners' behavioral data and their characteristics to understand learners' learning processes in a digital environment (Roll & Winne, 2015).

**Summary**

Overall, prior research highlights three key directions for advancing educational measurement in AI-assisted learning environments: analyzing temporal interaction dynamics, modeling multiple learning outcomes, and identifying behavioral-cognitive learner profiles. However, these dimensions are often studied separately. There remains a need for integrated research frameworks that combine temporal analysis, multi-outcome modeling, and learner profiling within AI tutor interaction environments. The present study addresses this gap by developing a prototype AI tutor agent framework that enables the systematic analysis of interaction logs generated during AI-assisted learning tasks.

### IV. Research Questions

**Research Question 1:** To what extent do students' early interaction patterns in AI-assisted tasks predict later task correctness and trust in the AI tutor?

**Research Question 2:** How do students' interaction patterns and response times change during AI-assisted tasks, and can these temporal dynamics predict later performance and trust in the AI tutor?

**Research Question 3:** Can we identify latent student profiles by combining behavioral metrics with cognitive indicators to predict responsiveness to AI feedback types?

### V. Experiment Setup

We propose a prototype AI tutor agent in this research aims to assist learners in completing tasks related to programming and data science. These tasks are related to general coding problems in which learners are asked to implement basic algorithms or complete incomplete code in solving problems like regression modeling and data clustering analysis using Python. During the interaction process, learners are required to submit their code attempts, and then they receive instructional feedback from the AI tutor. The feedback that the AI tutor offers includes answers, code, examples, explanations, and hints, which are intended to assist learners in solving problems correctly and conceptually. The interaction logs are the basis of the behavioral and cognitive indicators that are used in the analysis.

**Overall Experimental Design**

This paper presents an evaluation of the proposed adaptive prototype AI agent tutor framework through a training and deployment pipeline that mirrors how an intelligent tutoring system would monitor learner states, select pedagogically sound feedback, produce responses to those students, and record outcomes to be analyzed in the future. The whole process consists of four interconnected components: a tutor policy learning module, a tutor response generation module, an interaction logging module, and a series of analytic components for learner profiling and early warning system prediction.

More formally, a tutoring interaction can be viewed as a decision step in which a system observes a learner state vector $s_t$ and $a_t: s_t \in R^d$, $a_t \in A$, where the tutor policy is modeled as a neural network with parameter θ, such that $\pi_\theta(a_t \mid s_t)$ that outputs the probability of selecting an action $a_t$ given the learner's state $s_t$. The workflow starts with the training of the tutor policy on structured interaction data with warm-start labels that are close approximations to initial pedagogical decisions. The trained policy is then integrated into the tutor agent to facilitate adaptive feedback selection for learner interactions. The logged data are then leveraged to facilitate profiling and prediction, thus not only providing an instructional agent but an analytic framework for educational decision support.

**Implementation Environment**

The framework is implemented with the Python programming language, with the use of Python libraries, such as NumPy, Pandas, Scikit-Learn, PyTorch, etc. The implementation of the tutor policy network is based on the PyTorch library, while the implementation of the clustering and early warning modeling is based on the scikit-learn library. In order to ensure the reproducibility of the code, the random seeds were set to 42 for the random library, the numpy library, and the torch library. In addition, the code is designed to use the GPU computation capability if the CUDA library is available, or else the CPU computation capability is used.

**Tutoring Action Space**

The adaptive tutoring problem was formulated as a multi-class policy learning task in which the AI tutor chooses one feedback type from a predefined pedagogical action space. The AI tutor needs to decide each time what type of feedback to provide to the student. In this case, the model is designed to predict one of the six types of tutoring actions:
$A = \{hint, answer, explanation, example, quiz\ tip, code\ snippet\}$.

Given a learner state $s_t$, the policy neural network outputs logits $z$. Then, the probabilities over the actions are computed as $P(a = k \mid s_t) = \frac{exp(z_k)}{\sum_{j=1}^{K} exp(z_j)}$, where $s_t \in R^d$: the learner's current state including behavioral and contextual features (interaction indicators), $a \in A$: the tutoring action selected by the AI tutor, $k$: the index of specific feedback type within the action space, $z_k$: the score (logit) produced by the neural network for the $k$-th tutoring action before probability normalization, $K$: the total number of possible tutoring actions. In this study, $K = 6$ corresponds to the six feedback types as defined above. In this case, this formulation allows the AI tutor to capture the process of selecting feedback as a probabilistic decision process.

**Learner State Representation and Feature Construction**

For each interaction, the policy takes in a learner task state that is structured and contains learner baseline characteristics, task context, and short-term behavioral cues. The learner task state consists of state variables, and each state variable consists of learner prior knowledge, learner programming experience, learner motivation level, question type, question difficulty, response time, number of attempts, whether the learner requests a hint, and turn index. The learner state vector (representing the learner's observable attributes and behavioral indicators at time $t$) can be written as
$s_t = [pk_t,\ pe_t,\ m_t,\ d_t,\ rt_t,\ a_t,\ h_t,\ turn_t,\ q_t]$. $pk_t \in [0, 1]$ represents prior knowledge; $pe_t \in [0, 1]$ representing programming experience. $m_t \in [0, 1]$ represents motivation level. $d_t \in [0, 1]$ represents question difficulty. $rt_t \in [1, \infty)$ represents response time. $a_t \in \{1, 2, 3\}$ represents number of attempts. $h_t \in \{0, 1\}$ represents hint request (0 refers to no hint, and 1 refers to hint requested). $turn_t \in \{1,..., 9\}$ represents tutoring turn. $q_t \in \{0, 1\}^6$ represents question type (six categories: concept, debug, application, analysis, design, and optimization). Overall, there are 8 bahavioral and context features and 6 question types, so $s_t \in R^{14}$.

During the process of feature construction, the numerical features are standardized by the use of the expression $\bar{x} = \frac{x-\mu}{\sigma}$, which corresponds to the StandardScaler transformation. The variable representing the question types, $q_t$, being a categorical variable, is converted to a vector representation $q_t \in \{0, 1\}^K$, where $K = 6$ represents the six different types of questions. This vector representation of the question types is then concatenated to the standardized numerical variable.

**Warm-Start Training Data**

To begin with, the framework prepares a structured training set with 2,500 interactions to begin with. It consists of learner-task states along with learning outcomes. The probability of the learner providing the correct response can be represented as follows
$p_i = 0.55 + 0.25pk_i + 0.10pe_i + 0.05m_i - 0.20d_i - 0.10(a_i - 1) - 0.05h_i$, which is clipped to the interval $[0.05, 0.95]$: $p_i = clip(p_i, 0.05, 0.95)$. The probability is clipped to the interval $[0.05, 0.95]$ to avoid extreme probabilities and ensure realistic variability in the simulated learner responses.

Similarly, the correctness can be represented with the Bernoulli distribution: $y_i \sim Bernoulli(p_i)$. Other outcomes, such as quiz score, project completion, and improvement, are generated from bounded Gaussian distributions to reflect variation in learner performance and progress.

**Heuristic Policy Labeling**

A set of heuristic labeling rules are used to produce warm start training labels. The tutoring action is determined based on a rule-based approach based on the learner's current interaction state. Specifically, if the learner has attempted the problem at least three times and still answered the problem incorrectly (attempts >= 3 ∧ correctness = 0), the system will provide the learner with the answer. If the learner has asked for a hint and still answered the problem incorrectly (hint requested = 1 ∧ correctness = 0), the tutor will provide a hint. When the question type includes debugging, the system will provide a code snippet to the learner. When the question type includes analysis, the tutor will provide an explanation to the learner. When the learner answers a question quickly and incorrectly (response time < 10 ∧ correctness = 0), the tutor will provide a quiz tip to the learner. In other cases, the tutor will provide an example to the learner.

**Tutor Policy Model and Training Procedure**

The tutor policy model follows a feedforward network. Given a state vector $x$, the network computes $h_1 = ReLU(W_1 x + b_1)$, $h_2 = ReLU(W_2 h_1 + b_2)$, $z = W_3 h_2 + b_3$, where $z$ represents the output logits corresponding to the six tutoring actions. $x$ denotes the learner state vector. $W_1$, $W_2$, and $W_3$ denote the weight matrices of the neural network layers. $b_1$, $b_2$, and $b_3$ denote the bias vector. $h_1$ and $h_2$ denote the hidden layer representations. $ReLU(\cdot)$ denotes the rectified linear unit activation function.

- $h_1 = ReLU(W_1 x + b_1)$: The input learner state $x$ is linearly transformed by weights $W_1$ and bias $b_1$, then passed through a $ReLU$ activation to produce the first hidden representation $h_1$. It captures basic patterns from the learner state features through a nonlinear transformation.
- $h_2 = ReLU(W_2 h_1 + b_2)$: The first hidden representation $h_1$ is further transformed by a second layer with weights $W_2$ and bias $b_2$, and activated by $ReLU$ to obtain a higher-level representation $h_2$. It combines the patterns from the first layer to form a higher-level representation useful for predicting tutoring actions.
- $z = W_3 h_2 + b_3$: The second hidden representation $h_2$ is linearly mapped to the output vector $z$, which contains the logits corresponding to the six possible tutoring actions. It provides the logits that indicate the model's preference scores for the six possible tutoring actions.

Given the action space $A = \{hint, answer, explanation, example, quiz\ tip, code\ snippet\}$, the output vector $z$ contains six elements corresponding to the six tutoring actions, such that $z \in R^6$.

The model is trained using cross-entropy loss $L = -\sum_{i=1}^{N}\sum_{k=1}^{K} y_{ik} \log(\hat{p}_{ik})$, where $N = 2,500$ is the number of training samples, and $K = 6$ is the number of tutoring action classes. The Adam optimizer with a learning rate of $10^{-3}$ is used to optimize the model. The dataset is split into training and validation sets using an 80/20 stratified partition to preserve label balance. Inference follows a temperature-scaled softmax to convert logits to probabilities: $P(a = k \mid s) = \frac{\exp(\frac{z_k}{T})}{\sum_j \exp(\frac{z_j}{T})}$, where $T$ controls the level of stochasticity in the action selection process.

**Tutor Agent Deployment**

Once training is complete, the tutor policy is integrated into an end-to-end AI tutor agent that includes action selection, response generation, and logging. The learner's current behavioral and contextual features as $s_t$, an action is chosen based on the tutoring action $a_t \sim \pi_\theta(\cdot \mid s_t)$, where $\pi_\theta(\cdot \mid s_t)$ represents the probability distribution over possible tutoring actions given the learner state $s_t$. The feedback text generated by the AI tutor in response to the selected tutoring action at time $t$ is represented as $r_t$. The interaction consistent of learner state, the tutoring action, the response, and the outcome vector $o_t = (y_t, q_t, c_t, i_t, u_t, s_t, t_t)$ that includes correctness, quiz score, project completion, improvement, perceived usefulness, satisfaction, and trust. $o_t$ denotes the outcome vector at interaction step $t$; $y_t \in \{0, 1\}$ denotes the correctness; $q_t \in [0, 1]$ denotes the normalized quiz score; $c_t \in [0, 1]$ denotes the project completion rate; $i_t \in [0, 1]$ denotes learner improvement; $u_t \in [0, 1]$ denotes perceived usefulness; $s_t \in [0, 1]$ denotes learner satisfaction; $t_t \in [0, 1]$ denotes trust in the AI tutor.

**Reward Design**

To facilitate the multi-objective evaluation, the framework derives a composite reward from each interaction as follows:
$R = \alpha_p(0.5\ correctness + 0.5\ quiz\ score) + \alpha_i\ improvement + \alpha_t\ trust + \alpha_s\ satisfaction$,
where the weights are defined as $(\alpha_p, \alpha_i, \alpha_t, \alpha_s) = (0.45, 0.20, 0.20, 0.15)$. The reward is clipped to the interval $R \in [0, 1]$.

**Lerner Profiling**

To identify find the latent learner groups, the interaction data is aggregated at the student level and then K-means clustering is performed. With the feature vectors $x_i$ of learner $i$, the goal of the K-means clustering is to minimize the following objective: $min \sum_{i=1}^{N} ||x_i - \mu_{c_i}||^2$, where $c_i$ denotes the centroid of the cluster assigned to learner $i$, and $||x_i - \mu_{c_i}||^2$ represents the squared Euclidean distance between the learner feature vector and the corresponding cluster centroid.

**Early Warning Prediction**

The early warning component predicts the probability of task correctness based on early behavioral signals, enabling the system to identify learners who may require additional instructional support. The early warning part is based on the prediction of correctness of the task based on the early learner behavior cues using logistic regression. The probability of correctness is given by the equation $P(y = 1 \mid x) = \frac{1}{1+e^{-(\beta_0 + \beta^T x)}}$, where $x$ is the set of features such as response time, number of attempts, hint requests, turn index, prior knowledge, programming experience, level of motivation, and task difficulty.

**Experimental Scope**

In summary, the experiment has been designed with the aim of providing a proof-of-concept training and evaluation framework for adaptive AI tutors. The experiment setup facilitates evaluation at various levels: policy learning performance, functionality of the entire agent, and the value of the interactions for subsequent learner analysis. Although the proposed experiment is designed to be an initial development-stage framework, it has already captured the basic structure of a trainable AI tutor: state representation, policy learning, action generation, reward modeling, and learner analysis.

## VI. Methodology

### (VI.1) Research Question 1

**Longitudinal Interaction Data Generation**

To examine the relationship between early interaction behaviors and subsequent learning outcomes and trust in the AI tutor, a longitudinal interaction dataset was created by using the implemented AI tutor prototype. After the warm-start policy training, the tutor agent was used to interact with 40 simulated students, and each student completed 8 sequential turns of interaction, generating a total of 320 temporally ordered interaction records.

More specifically, each interaction at time step $t$ for student $i$ can be defined as follows $x_{it} = (s_i, b_{it}, o_{it})$, where $s_i$ refers to the learner's baseline attributes, such as their prior knowledge, programming skill level, and motivation, $b_{it}$ refers to the learner's behavioral attributes, such as response time, number of attempts, hint requests, and turn index, and $o_{it}$ refers to the learner's observed outcomes, such as correctness, quiz score, trust, and satisfaction. The repeated measurements of each learner's interaction yield the dataset, which includes both cross-sectional and temporal variations in learner behavior.

**Temporal Behavior Analysis**

To identify general temporal patterns in AI-assisted learning, interaction logs are first aggregated by a tutoring turn index. For each turn $t$, the average behavioral and outcome variables are calculated among students: $\bar{x}_t = \frac{1}{N} \sum_{i=1}^{N} x_{it}$. $x_{it}$ denotes the value of that indicator for student $i$ at turn $t$; $N$ denotes the number of students; $t$ denotes the tutoring turn index. The temporal indicators to be analyzed are as follows: average response time, average hit request rate, average number of attempts, average correctness of tasks, and average trust in AI tutor. These aggregated trajectories can be used to visualize the evolution of student behavior over multiple interactions with the AI tutor, such as reduced response times or increased correctness over multiple turns.

**Early Interaction Feature Construction**

In order to evaluate whether early behavioral trends are predictive of later outcomes, a student-level dataset is constructed based on features extracted from each student's early interaction window. In the current experiment, the early interaction window is defined as the initial three turns of tutoring, $E_i = \{x_{i1}, x_{i2}, x_{i3}\}$.

Summary statistics are then obtained for this early window. These include the average response time, average number of attempts, hint request rate, correctness rate, and average trust level.

In addition, a set of temporal trend features are obtained by fitting a linear function to the early turns. These include the response time slope, which reflects whether the response time increases or decreases over the early interaction phase. A corresponding slope for hint usage is also obtained. The resulting student-level feature vector is $z_i = (pk_i, pe_i, m_i, \bar{rt}_i, \bar{a}_i, \bar{h}_i, \bar{c}_i, \bar{trust}_i, \beta_i^{(rt)}, \beta_i^{(hint)})$. $pk_i$ refers to the prior knowledge of learner $i$. $pe_i$ refers to the programming experience of learner $i$. $m_i$ refers to the motivation level of learner $i$. $rt_i$ refers to the average response time of learner $i$ during the early interaction window. $a_i$ refers to the average number of attempts made by learner $i$ during the early interaction window. $h_i$ refers to the hint request rate of learner $i$ during the early interaction window. $c_i$ refers to the correctness rate of learner $i$ during the early interaction window. $trust_i$ is the average trust level reported by learner $i$ during the early interaction window. $\beta_i^{(rt)}$ is the slope of the linear regression fitted to the response time values across the early interaction turns for learner $i$, which shows whether the response time for learner $i$ increases or decreases over time. $\beta_i^{(hint)}$ is the slope of the linear regression fitted to the hint usage across the early interaction turns, which shows the temporal and trend of hint requests during the early interaction phase.

**Predicting Final Learning Outcomes**

In order to determine if early interaction patterns are indicative of later task performance, a logistic regression model can be used to predict each student's final task correctness $P(y_i = 1|z_i) = \frac{1}{1+e^{-(\beta_0 + \beta^T z_i)}}$, where $z_i$ is the feature vector for early interactions and $y_i$ is the final correctness. Model performance can be assessed by prediction accuracy.

**Predicting Trust in the AI Tutor**

In addition to task correctness, the analysis also looks into whether the early behavioral signals can forecast the level of trust that the students will have in the AI tutor later on. Because trust is measured as a continuous variable, a linear regression model is used: $trust_i = \gamma_0 + \gamma^T z_i + \epsilon_i$. Model performance is evaluated using the root mean squared error ($RMSE$) and the coefficient of determination $R^2$.

**(VI.2) Research Question 2**

**AI Tutor Policy Model**

To simulate the interaction between adaptive AI-assisted tutoring and the learner, a tutor policy model was first learned using a synthetic dataset of tutoring instances. Each instance contained learner attributes and behavioral features, such as knowledge, programming skill level, motivation level, question types, difficulty levels, response times, number of attempts, hint requests, and turn index. A feedforward neural network model was learned to predict the tutor's feedback type, such as hint, explanation, and example, from the features. This learned policy model was then used to select the feedback provided by the AI tutor.

**Longitudinal Interaction Simulation**

Using the trained policy model, longitudinal tutoring sessions were simulated for $N = 40$ students across $T = 8$ interaction turns. At each turn $t$, behavioral variables including response time $RT_{it}$, number of attempts $A_{it}$, hint requests $H_{it}$, correctness $C_{it}$, and trust $T_{it}$ were generated based on learner characteristics, task difficulty, and turn progression. Turn-level summaries were computed as averages across students, for example $\bar{RT}_t = \frac{1}{N_t} \sum_{i=1}^{N_t} RT_{it}$, where $N_t$ denotes the number of students observed at turn $t$. These summaries capture the overall temporal evolution of interaction behavior during AI-assisted tasks.

**Temporal Feature Extraction and Predictive Modeling**

To examine the temporal dynamics at the student level, behavioral features were extracted from the interaction trajectory. Early stage averages were calculated for the first $k = 3$ turns, and temporal trends were summarized by linear slopes over the turns. Early-to-late behavioral change was defined as $\Delta_i(y) = \bar{y}_{i\ (late)} - \bar{y}_{i\ (early)}$, where $y$ represents behavioral variables such as response time, attempts, hint use, correctness, or trust. These temporal features are then used to train a model to predict future outcomes. Logistic regression was used to predict the final correctness of the task $P(Y_i = 1 \mid X_i) = \frac{1}{1+e^{-(\beta_0 + X_i^\top \beta)}}$, where $X_i$ denotes the vector of temporal and baseline features for the student $i$. Linear regression was used to predict final trust in the AI tutor $T_i^{final} = \beta_0 + X_i^\top \beta + \epsilon_i$.

These models will allow the analysis to investigate if longitudinal interaction patterns in the tutoring sessions are predictive of the students' performance/trust outcomes.

### (VI.3) Research Question 3

An analysis pipeline for student-level analysis has been developed to investigate if learner profiles could be identified using behavioral and cognitive features, and if this could be used to predict the response to various types of AI feedback.

**Student-Level Feature Construction**

First, the interaction logs were aggregated at the student level. Student-level aggregates were computed as the combination of the student's baseline cognitive features, which included their knowledge level, programming experience, and motivation level. In addition, the student's behavioral features were aggregated, which included their average response time, average number of attempts, hint request rate, correctness, trust, satisfaction, and reward. Furthermore, the student's temporal features were extracted based on the slope of the student's turns over the tutoring sessions with respect to their response time, number of attempts, hint request, correctness, and trust. In general, for a variable $y_{it}$ measured for student $i$ at turn $t$, the student-level mean was defined as $\bar{y}_i = \frac{1}{T_i} \sum_{t=1}^{T_i} y_{it}$, where $T_i$ is the total number of interactions for student $i$.

**Latent Profile Identification**

The next step was to standardize the student-level feature matrix and cluster it. This clustering was done using the K-means method. In this method, the objective was to identify the latent profiles of the students based on similar cognitive and behavioral characteristics. In the K-means method, the clustering of the students is done by minimizing the variation: $min_{C_1,...,C_k} \sum_{k=1}^{K} \sum_{X_i \subseteq C_k} ||X_i - \mu_k||^2$. $X_i$ is the feature vector for student $i$. $\mu_k$ is the centroid of cluster $k$.

**Feedback Responsiveness Modeling**

To determine this, a responsiveness score was formulated for a student i to a given type of feedback $f$ as the difference between the student's mean reward under this type of feedback $f$ and the student's overall mean reward $R_{i,f} = \bar{r}_{i,f} - \bar{r}_i$. A positive value of this type means that this type of feedback $f$ was more effective than usual for a given student $i$. The type of feedback $f$ with the highest value of this type was used to define a given student's most effective type of feedback strategy. Two predictive models were also used. A multinomial logistic regression model was used to forecast a given student's most responsive type of feedback, whereas a linear regression model was used to forecast this type's best value of a responsiveness score. The results obtained from this analysis were used to test whether it was possible to identify a latent learner profile.

## VII. Result

### (VII.1) Research Question 1

**Predictive Value of Early Interaction Patterns**

To test the effectiveness of the early interaction as a predictor of the final learning outcomes and the level of trust in the AI tutor, two predictive models were developed. These included a logistic regression model to predict the final level of correctness for the tasks and a linear regression model to predict the final level of trust in the AI tutor.

**Temporal Interaction Patterns Across Tutoring Turns**

The interaction logs show that there are several systemic behavior trends in the process of interacting with the tutor. The response time decreases steadily in the process of interacting with the tutor (Figure 1). From the entire process of simulated interaction, it is clear that the average response time for the students decreases steadily from earlier to later turns in the process. This implies that as the students interact with the AI tutor, they become more efficient in responding to the tasks, possibly due to better familiarity with the structure of the tasks and the instructional feedback from the tutor. The hint request rates show a steady decline in the process of interacting with the tutor (Figure 2). From the entire process of interacting with the tutor, it is clear that the students request hints more often in earlier turns, but as the process advances, the request for hints diminishes steadily.

**Figure 1**
*Response Time Across Turns*

**Figure 2**
*Hint Requests Across Turns*

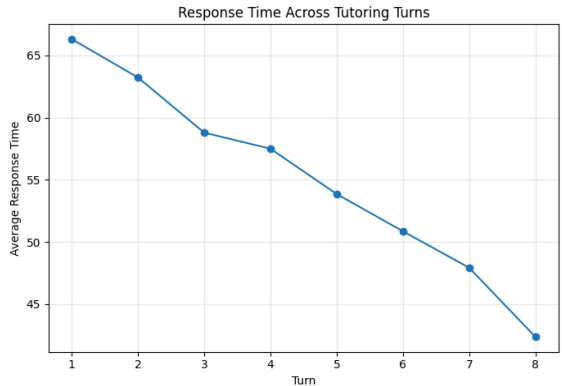
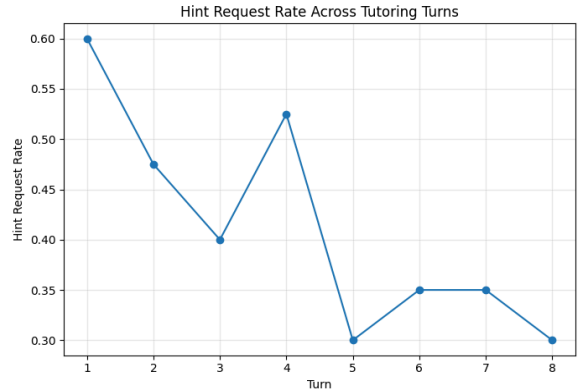

The correctness level in completing tasks rises significantly across the tutoring turns (Figure 3). The average correctness level rises over time in the tutoring sequence and reaches a high level in the last turns. This rising pattern indicates that the tutoring process facilitates progressive learning and enhanced task mastery over time. The trust level of the students in the AI tutor rises across the tutoring turns (Figure 4). The average trust level rises over time, showing that the trust level of the students in the AI tutor rises over time. This rising pattern indicates that as the students achieve success in completing tasks and receive effective assistance, their trust level in the AI tutor rises gradually.

**Figure 3**
*Correctness Across Turns*

**Figure 4**
*Trust Across Turns*

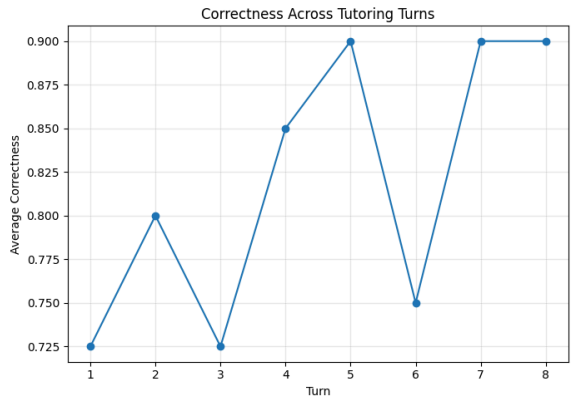
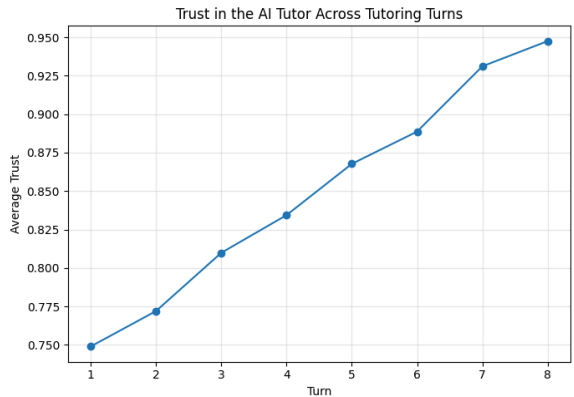

Taken together, these temporal patterns indicate that students gradually respond faster, request fewer hints, achieve higher correctness rates, and develop greater trust in the tutor over the course of the interaction sequence.

**Prediction of Final Task Correctness**

In order to test whether early interaction patterns are predictive of later task performance, a logistic regression model was employed based on the features learned in the first three turns of the tutoring session. These features include learner baseline factors (prior knowledge, programming experience, and level of motivation), early learner statistics (mean response time, mean number of attempts, early hint rate, correctness rate, and mean trust), and early trends (response time slope and early hint slope). The logistic regression model had an accuracy of 0.925 on the training data, implying that early learner behavior does indeed hold predictive value in relation to later correctness in the simulated experimental environment. From the logistic regression model, it can be seen that several early learner behavior indicators are predictive of later success. The early hint slope, in particular, had the strongest negative coefficient of -0.952, implying that learners showing an increased tendency to use early hints in the tutoring session are less likely to be correct in the end. Prior knowledge and early correctness rate also had negative coefficients, although they were less significant in comparison to the early hint slope. Motivation, early hint rate, and programming experience had positive coefficients, although they were less significant in comparison to the early mean attempts, prior knowledge, correctness rate, and early hint slope. The early mean attempts, in particular, had the highest positive coefficient of 0.575, implying that, in the simulated environment, the relationship between early mean attempts and later correctness is more nuanced than one of absolute deficits.

**Prediction of Final Trust in the AI Tutor**

In addition to the correctness of the tasks, a linear regression model was employed to predict the final trust level of the students based on the early interaction features. A training $RMSE$ of 0.0275 and $R^2$ 0.8352 were achieved by the model. This revealed that the early behavioral and attitudinal characteristics of the students explained a significant portion of the final trust level.

The results showed that the predictor with the highest correlation with the final trust level was the early trust level itself, with a correlation of 0.669. This revealed that the students who trusted the AI tutor in the early stages would also trust it in the final stages. Moreover, the motivation level of the students (0.036) and the prior knowledge of the students (0.026) had a positive correlation with the final trust level. On the other hand, the programming experiences of the students (-0.079), correctness rate in the early stages (-0.080), early stages' average attempts (-0.025), and hint slope in the early stages (-0.046) had negative correlations. Overall, the results revealed that the early interaction characteristics of the students were useful in predicting the changes in the trust level of the students for the AI tutor.

**Summary of Research Question 1**

Overall, the results suggest that early interaction patterns in AI-assisted learning environments contain informative signals that can be used to predict later task correctness as well as students' trust in the AI tutor. More specifically, the logistic regression analysis found that early interaction features could be used to predict final task correctness with high training accuracy (0.925), while the linear regression analysis found that early interaction features explained a large proportion of the variance in final trust ($R^2$ = 0.8352). Overall, the results of this study suggest that early interaction signals have the potential to support the early identification of students who may benefit from additional instruction in AI-assisted learning systems. Of course, the current results should be interpreted as a demonstration of the methodological feasibility of the simulated interaction framework that is fully replicable.

**(VII.2)   Research Question 2**

**Temporal Changes in Interaction Patterns**

The turn-level summaries indicate clear temporal changes in students' behavioral patterns during AI-assisted tasks. For each tutoring turn $t$, the average response time can be expressed as

$\bar{RT}_t = \frac{1}{N_t} \sum_{i=1}^{N_t} RT_{it}$, where $RT_{it}$ denotes the response time of student $i$ at turn $t$, and $N_t$ is the number of students observed at that turn. Across tutoring turns, $\bar{RT}_t$ generally decreased, suggesting that students became more efficient as they progressed through repeated interactions with the AI tutor. Similarly, help-seeking behaviors, reflected in hint requests and the number of attempts, tended to decline over time. These turn-level quantities can be summarized as $\bar{C}_t = \frac{1}{N_t} \sum_{i=1}^{N_t} C_{it}$ and $\bar{T}_t = \frac{1}{N_t} \sum_{i=1}^{N_t} T_{it}$, where $C_{it} \in \{0, 1\}$ indicates correctness and $T_{it}$ denotes trust. The increase in the values of $\bar{C}_t$ and $\bar{T}_t$ indicates that the repeated interaction with the tutor was related to an improvement in the success of the task and the confidence in the AI tutor. Overall, the patterns indicate a coherent longitudinal learning process in which the student transitions from less efficient and more assistance-dependent behavior towards more efficiency, accuracy, and confidence.

Further insights can be gained from the response time trajectories of the students (Figure 5). Overall, the response time was declining in all six task categories, namely analysis, application, concept, debug, design, and optimization, from Turn 1 to Turn 8. This indicates that the temporal efficiency gain was not restricted to a single task format. Although minor variations in the response time are seen in the intermediate turns, particularly around Turns 3-4, the overall trend is declining in all the question types. This suggests that the students are becoming progressively more efficient not only in interacting with the AI tutor in general but also in different types of cognitive engagement.

**Figure 5**

*Response-Time Dynamics by Question Type*

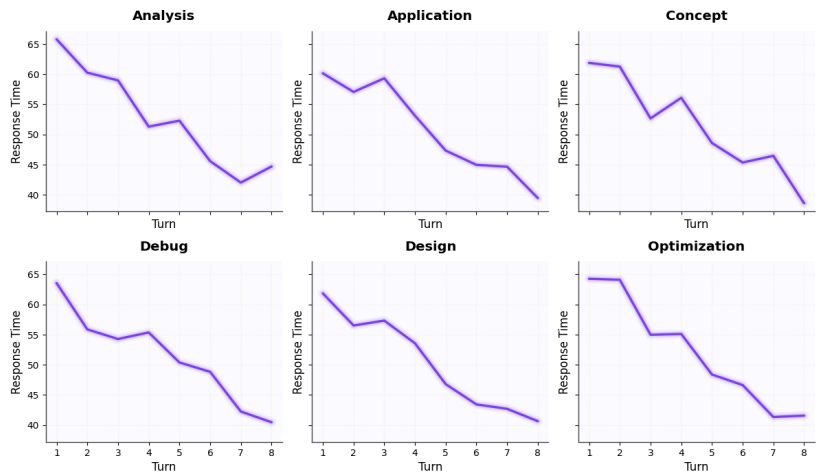

In addition, there are some task types that seem to reduce more dramatically than others. For example, the application, concept, and optimization types had higher response time values initially for the early turns and reduced more dramatically for the final turns. This might suggest that these types are more cognitively demanding initially but become more manageable with repeated interaction with the AI tutor. The debug and design types had a similar trend with a decrease over the turns. The analysis types had a similar trend with a minor spike for the final turn. However, the overall picture provided by this figure confirms that the reduction of response time is a strong temporal trend for all instructional contexts simulated. This supports the interpretation that the students are building familiarity with the tasks, procedural fluency, and interaction efficiency over time.

The overall trend for the response time over the turns, based on the distribution of response time for the tutoring turns with a ridgeline plot (Figure 6), shows a gradual reduce in response time across turns. This suggests that the students are responding faster over time. The earlier turns tend to have higher and more variable response times, while the later turns tend to be faster and more concentrated. Further evidence of the temporal process is found in the relationship between response time and trust across tutoring turns. The Hexbin map (Figure 7) displays the relationship between response time and trust. In this figure, the densest interaction zone is centered at the middle range of response time and trust. Thus, the majority of student interactions occurred at a stable zone. What is more important is that the trajectory of interaction over time appears as a progression from the lower-right area of the plot, where the value of response time is high and the value of trust is low, to the upper-left area of the plot, where the value of response time is low and the value of trust is high. Therefore, the process of interaction between the student and the AI system over time is one where the student is not only getting more efficient at the AI-assisted tasks but is also becoming more trusting of the tutor.

| **Figure 6** | **Figure 7** |
|---|---|
| *Ridgeline Plot of Response Time Across Tutoring Turns* | *Hexbin Phase Map of Response Time and Trust* |

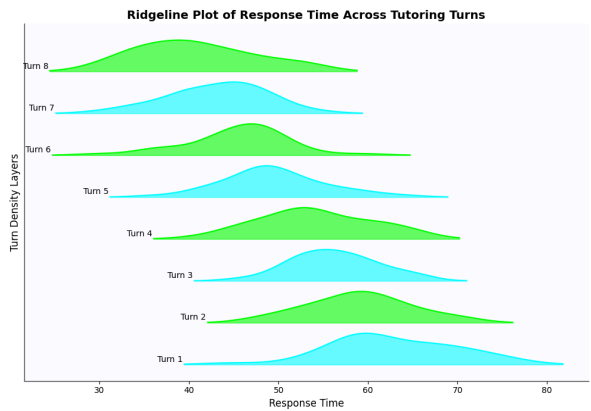
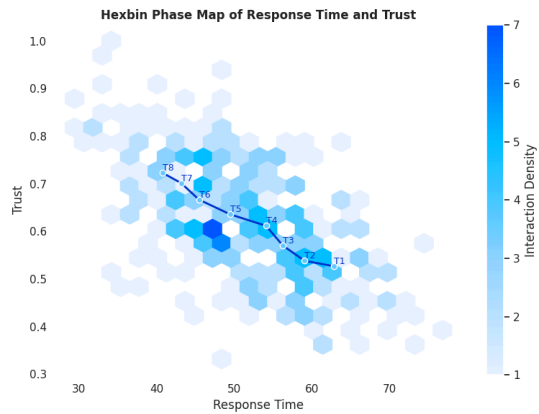

Student-level features further support this interpretation. For each student $i$, the temporal slope of a behavioral variable $y_{it}$ over turns was summarized as the coefficient from a simple linear trend, and early-to-late change was defined as $\Delta_i(y) = \bar{y}_{i\ (late)} - \bar{y}_{i\ (early)}$. Using these summaries, it can be noted that some students had negative response time slopes, which indicated that their response time reduced over the turns. Additionally, some students had negative late-minus-early changes in response time, attempts, and hint use, which indicated a reduction in effort burden and support need. However, some students had positive changes in correctness and trust from early to late interaction phases.

**Predicting Later Performance from Temporal Dynamics**

To evaluate whether these temporal processes were related to subsequent learning outcomes, a logistic regression model was developed to predict final task correctness using student-level temporal features. More formally, the probability of final correctness for a student $i$ was modeled as $Pr(Y_i = 1 | X_i) = \frac{1}{1+e^{-(\beta_0 + X_i^T \beta)}}$, where $Y_i$ represents final correctness, and $X_i$ represents the vector of temporal and baseline features of the student. The model achieved a training accuracy of 0.850, showing that the temporal features extracted contained significant predictive information for the success of the tasks.

The strongest positive predictors of final correctness were improvement in correctness over time (1.4187), prior knowledge (0.8875), and motivation level (0.5964). This indicates that the more improvement there was in correctness over time, the higher the prior knowledge of the student, and the higher the level of motivation, the higher the probability of correctness in the final turn. Positive predictors of correctness in the final turn were correctness slope, early hint rate, early mean attempts, and early mean trust.

The results show that there are certain negative predictors of correctness in the final turn as well. The strongest negative predictors of correctness in the final turn are attempt change (-0.6012) and response time slope (-0.5815). This indicates that the more attempts the students needed in the early part of the session, the less likely they were to be correct in the final turn. The faster the students responded, the less clear the pattern was, and the less likely they were to be correct in the final turn. Programming experience was a negative predictor of correctness in the final turn as well, with a coefficient of -0.3014. The simulated nature of the data may have influenced the results. However, the results show that the characteristics of the students and the changes in the behavior of the students over time both play a part in the prediction of correctness in the final turn.

**Predicting Final Trust from Temporal Dynamics**

A linear regression model was then trained to predict final trust in the AI tutor using the same set of temporal features. The prediction equation can be written as $T_i^{final} = \alpha_0 + X_i^T \alpha + \varepsilon_i$, where $T_i^{final}$ denotes the student $i$'s final trust score, $X_i$ is the temporal feature vector, and $\varepsilon_i$ is the residual term. The model produced a training $RMSE$ of 0.0221 and a training $R^2$ of 0.9593, indicating that the temporal features explained a very large proportion of the variance in final trust within the simulated environment.

For evaluation, the prediction error was quantified using the root mean squared error,

$RMSE = \sqrt{\frac{1}{n} \sum_{i=1}^{n} (T_i^{final} - \hat{T}_i^{final})^2}$ , and explanatory strength was summarized by

$R^2 = 1 - \frac{\sum_{i=1}^{n}(T_i^{final} - \hat{T}_i^{final})^2}{\sum_{i=1}^{n}(T_i^{final} - \bar{T}_i^{final})^2}$. The strongest positive predictor of final trust was trust slope (coefficient = 15.4440), followed by early mean trust (coefficient = 0.8262). This indicates that students who had high initial trust levels and a smooth increase in trust levels throughout the conversation are more likely to end up with high trust levels in the AI tutor. Additional positive factors included hint slope (0.1879), early mean attempts (0.1394), attempt change late minus early (0.0947), and correctness change late minus early (0.0793), although these factors are much weaker than those of trust.

The negative values for a number of features included trust change late minus early (-1.5768), trust variability (-0.4035), correctness slope (-0.3323), early hint rate (-0.1480), and hint change late minus early (-0.1197). These negative values should be considered with caution, since some of these features are highly related to one another and might be a sign of multicollinearity for the regression model. However, the results provided by the model are quite clear on one thing: the patterns over time, particularly with respect to trust, are highly indicative for the prediction of the students' final trust levels.

**Summary of Research Question 2**

In the process, the pattern of students' interactions with the AI tutor also varied systematically with time. Specifically, it was observed that response times, attempts, and hint-seeking reduced over time, whereas correctness and trust increased with time. These observations are indicative of the fact that students were becoming more efficient, accurate, and trusting of the AI tutor with time.

Furthermore, it has also been found that the temporal patterns derived from the students' interactions with the AI tutor were strongly correlated with the outcomes. Specifically, it has been found that the logistic regression analysis revealed that the temporal features were significant predictors of the final correctness of the students' task, whereas the linear regression analysis revealed that the temporal features were highly correlated with the final trust levels in the AI tutor.

**(VII.3) Research Question 3**

**Latent Student Profiles and Feedback Responsiveness**

To investigate whether it's possible to identify student profile types based on their behavioral and cognitive characteristics, a clustering analysis on the features of individual students based on their interaction logs was performed. The dataset contains 320 interactions from 40 students, with five feedback types provided to the students: answer, code snippet, example, explanation, and hint.

**Latent Learner Profiles**

Four latent learner profiles were identified based on aggregated interaction features. Let the feature vector for the student $i$ be $x_i = (prior_i, prog_i, correct_i, trust_i, hint_i, attempt_i)$ where these variables represent prior knowledge, programming experience, correctness rate, trust level, hint usage, and attempts. Students are grouped by clustering $c_i = arg\ min_k ||x_i - \mu_k||^2$, where $\mu_k$ denotes the center of the learner profile $k$.

The resulting profile shows a clear difference. Profile 3, which includes 14 students, has the best performance with the highest correctness (0.661) and trust (0.711). Profile 2, with 11 students, has the worst correctness (0.364) but the best hint usage (0.568). Profiles 0 and 1 are in the middle with moderate performance. Overall, the response time slopes for all the profiles are negative, indicating that the response time decreases as the tutoring session proceeds.

**Predicting Feedback Types**

To check whether learner characteristics could be used to predict the type of feedback that the AI tutor might provide, a multi-class classification model was trained on the student features. The classification model resulted in a training accuracy of 0.55. Thus, the student's behavioral and cognitive characteristics were found to provide moderate predictive information regarding the type of feedback that might be provided. In addition, the type of feedback that the student might receive or that might be the best type of feedback for the student were found to be code snippet feedback, which resulted in the highest recall at 0.78. Hint feedback was found to be difficult to predict, mainly because it occurred infrequently.

To further quantify student responsiveness to AI tutor feedback, a regression model was used to predict a responsiveness score $R_i$. The model can be written as $R_i = \beta_0 + \sum_k \beta_k X_{ik} + \epsilon_i$, where $X_{ik}$ represents the set of student-level predictors and $\beta_k$ denotes the corresponding regression coefficients. The regression model achieved $RMSE = 0.308$ and $R^2 = 0.8426$, indicating that the selected features explained a large proportion of the mean improvement $\beta = 0.847$ and mean correctness $\beta = 0.451$. These findings suggest that students who show increasing trust in the AI tutor and stronger learning gains tend to respond more positively to AI-generated feedback.

A simple interpretation of responsiveness can also be expressed as a combination of performance improvement and trust change, such that $R_i = \Delta Correctness_i + \lambda \Delta Trust_i$, where $\Delta Correctness_i$ represents improvement in task performance, $\Delta Trust_i$ represents change in trust toward the AI tutor, and $\lambda$ is a weighting factor. Under this formulation, responsiveness reflects both cognitive improvement and affective adaptation during the tutoring process.

**Profile-Specific Feedback Preferences**

Further differences were found across the learner groups based on the feedback patterns that were specific to the learner profiles (Figure 8). Specifically, the most frequently associated feedback type with the learner profile 0 was explanation-based feedback. Therefore, learners with this profile may require more explanation-based feedback. For the learner profile 1, the feedback types were more evenly distributed between example-based and explanation-based feedback. For the learner profile 2, the feedback type that was most strongly associated with this learner group was example-based feedback. Therefore, this type of feedback may be more beneficial to learners with lower correctness and higher help-seeking behavior. On the other hand, the learner profile 3 showed the strongest association with the code snippet-based feedback type, which is consistent with their performance in the programming-related tasks and overall learning outcomes. Latent learner profiles may be identified based on the learners' behavioral and cognitive indicators.

The radar chart (Figure 9) compares the behavioral and performance characteristics of the four learner profile types. The profile with the best performance is profile 3. This profile has the highest values for motivation, correctness, improvement, trust, and reward. This means that these learners achieve the best results in learning and are more positive when interacting with the AI tutor. Profile 2 has high values for response time, attempts, and hints. This profile shows that these learners are more reliant on the tutor and need more attempts to finish the learning process. Profile 1 shows the best values for prior knowledge. However, profile 0 has relatively low values for all of the characteristics. This means that these learners have the best prior knowledge, but their engagement with the learning process is limited. Profile 1 shows relatively moderate values for most characteristics. This means that this profile does not have extreme values for any characteristic. Overall, the radar chart shows that there are differences in learner behavior. This means that learners interact with the AI tutor differently based on their prior knowledge, engagement, and learning progress.

**Figure 8**

*PCA Projection of Learner Interactions and Feedback*

**Figure 9**

*Radar Comparison of Latent Learner Profiles*

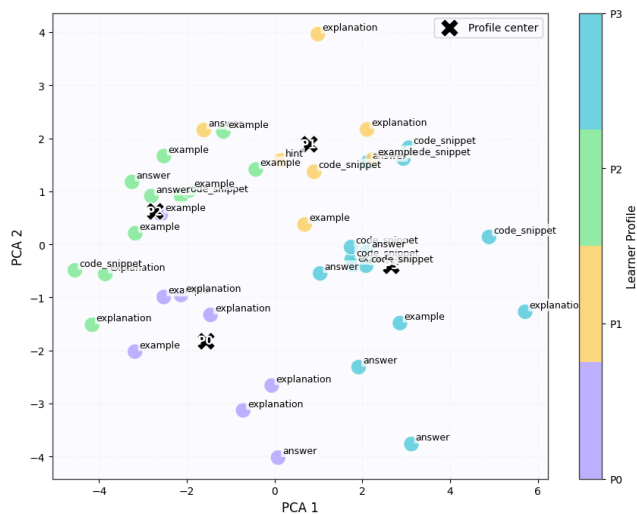
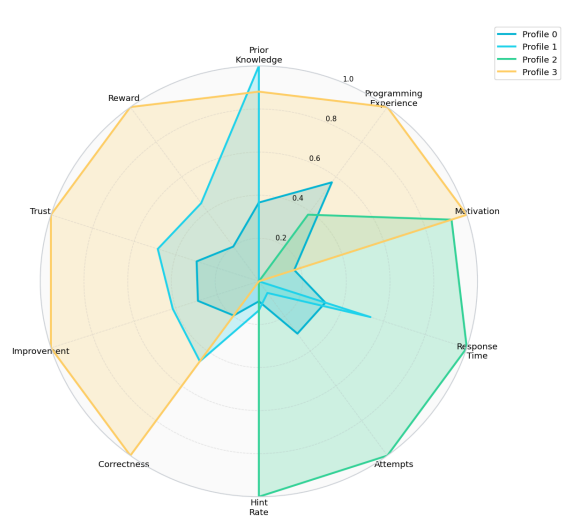

## VIII. Conclusion

This study aimed to investigate the relationship between student interactions in an AI-assisted environment and later outcomes, temporal learning trends, and learner differences. For Research Question 1 that focusing on Early Interaction Signals and Later Outcomes, early student interactions can be used to predict later outcomes. In this study, response time, attempts, hint use, correctness, and early trust were significant features that can be used to predict later outcomes. For Research Question 2 that focusing on Temporal Interaction Dynamics and Outcome Prediction, there are significant changes in student interactions with the AI tutor over time. These features were significant in predicting later outcomes. For Research Question 3 Latent Learner Profiles and Feedback Responsiveness, latent learner profiles are significant in explaining learner differences in response to the AI tutor's feedback. Learner profiles can be used to explain learner differences in response to the AI tutor's feedback. Overall, this study demonstrated the importance of early interactions, temporal learning trends, and learner differences in understanding the learning process in an AI-assisted environment. It demonstrated the importance of the reproducible framework in understanding adaptive learning processes. It can be concluded that this study has laid the foundation for future studies using learning data.

## IX. Discussion

The results obtained in this study are largely in line with the results obtained by other researchers on the use of AI-assisted tutoring, educational measurement, temporal learning analytics, multi-outcome assessment, and behavioral-cognitive learner profiling.

Firstly, the results obtained in this study are largely in line with the results obtained by other researchers on the use of AI-assisted tutoring and intelligent tutoring systems. In this regard, earlier researchers had established that the use of intelligent tutoring systems improves learning by providing timely guidance and step-by-step tutoring (VanLEHN, 2011; Chrysafiadi et al., 2022). Similarly, the results obtained for this study indicated that the correctness and trust levels improved over time, while the response time and dependence on hints reduced. These results are largely in line with the results obtained by other researchers on the use of intelligent tutoring systems, which improve learning by providing timely guidance and step-by-step tutoring. In this respect, the results obtained for this study are largely in line with the results obtained by other researchers on the use of intelligent tutoring systems. Therefore, the results obtained for this study are largely in line with the results obtained by other researchers on the use of intelligent tutoring systems.

Secondly, the results obtained for this study are largely in line with the results obtained by other researchers on the use of temporal interaction dynamics for intelligent tutoring systems and learning analytics. In this regard, earlier researchers had established that process data, such as response time, help-seeking behavior, and interaction sequences, can be used to identify learning patterns and predict future outcomes (Koedinger et al., 2015; Aleven et al., 2003; Ferguson, 2012; Roll & Winne, 2015). The results obtained for this study are largely in line with the results obtained by other researchers on the use of process data for learning analytics. The results obtained for Research Questions 1 and 2 indicated that early behavioral indicators, such as response time, hint usage, attempts, correctness, and trust, are predictive of later task correctness and trust with the AI tutor. Similarly, the results obtained for this study indicated that temporal features, such as slopes and early-to-late changes, are also informative for predicting later performance and trust.

Third, the research is consistent with the literature that highlights the value of considering multiple outcomes in the context of AI-supported learning. Previous research has indicated that the effectiveness of a digital learning platform should not be based solely on the achievement or correctness of the platform but should also be based on perception-based and process-based outcomes, such as satisfaction and usefulness, as well as trust (Tempelaar et al., 2015; Henrie et al., 2015). In this research, the effectiveness of the AI-supported platform is based on the achievement or correctness of the final answer as well as the perception-based outcome, trust. As the analysis demonstrated, trust is not only a significant outcome but is one that varies over time and is predictable from early and temporal interaction features, thereby supporting the literature that the effectiveness of educational measurement should be based on multiple outcomes rather than performance.

Fourth, the results are consistent with the findings of previous studies on learner profiling from the behavioral-cognitive perspective. In fact, it has been established in the literature that clustering analysis can be used to uncover hidden learner groups with varying levels of engagement, learning strategies, and performance (Romero & Ventura, 2010; Tempelaar et al., 2015). In this study, four latent learner profiles were revealed using aggregate learner features from the behavioral and cognitive aspects, which varied significantly in terms of correctness, trust, hint, attempts, and feedback. This is consistent with the literature that learner heterogeneity is an important aspect of digital learning environments, with evidence that learner features combined with behavioral features can improve understanding of learner interactions with AI tutors.

More specifically, the profile-specific feedback results extend this body of research in that they indicate that different learner groups might benefit from different levels of instructional support. For example, students with lower correctness and higher levels of help-seeking tended to be more associated with example-based feedback, whereas stronger-performing students tended to be more associated with code-snippet feedback. This is all in line with the general body of research on adaptive instruction in general, in that it suggests that effectiveness of feedback will depend on learner needs and preparedness. Therefore, this current research not only extends previous research on learner profiling but also does so in a way that integrates a more adaptive perspective on feedback within an AI tutor setting.

However, this current research also addresses a gap in the literature as identified in this review. As mentioned above, temporal interaction analysis, multi-outcome modeling, and learner profiling are all typically separate topics of research. The current framework integrates all of these aspects within a single AI tutor setting. By doing so, this current research provides a more integrated framework for measuring AI-assisted learning. Therefore, in addition to being aligned with previous research in this area, this current research also extends previous research in a way that integrates all of these separate aspects of AI-assisted learning within a single setting.

However, the alignment with the literature must also be taken into account appropriately. This is because the studies discussed in the literature review were conducted using real student data, whereas the current study used a simulated but reproducible interaction framework. Therefore, the results of the current study must be taken as a methodological proof of concept rather than empirical evidence for the results of the literature. However, the fact that the results of the current study align with the literature from a theoretical point of view speaks well for the plausibility and practical application of the proposed framework.

The discussion above indicates that the results of the current study align well with the literature review. This is because the current study provides empirical evidence for the importance of process data, temporal interaction modeling, multi-dimensional outcome evaluation, and learner profiling, as discussed in the literature. However, the current study provides a new contribution to the literature by combining the different perspectives of the literature into a single framework for the reproducibility of the results.

## X. Contribution & Implication

This research makes several contributions and holds significant implications for research on AI-Assisted Tutoring. Firstly, it reveals that early interaction behaviors, such as response time, attempts, hint usage, correctness, and early trust, are effective in predicting later task performance and student trust in the AI tutor. Secondly, it proves that student interaction behaviors vary over time, and that they tend to be faster, require fewer attempts, achieve higher correctness, and build trust in the AI tutor over time. Thirdly, it reveals that it is possible to classify students into distinct and meaningful latent profiles based on their behavioral and cognitive features, and that these profiles are effective in predicting differences in feedback responsiveness and learning. The research demonstrates that an AI tutoring systems have benefit in early prediction, temporal interaction, and profiling in developing effective adaptive and personalized tutoring strategies. It offers an effective research framework for investigating adaptive learning processes and is an important starting point for future research on educational data.

## XI. Limitation

The tutoring dialogues studied in this research were not based on a fully developed real-world AI tutor that has already been used in real-world educational settings. Instead, they were based on a prototype AI agent tutor developed as part of this research. In this respect, the prototype used in this research is not only fully developed but is also capable of generating adaptive feedback. In this respect, the prototype is fully capable of generating adaptive feedback. However, the prototype's capability to select the type of feedback to be used is limited to the action space that is defined as part of this research. In this respect, this means that the type of feedback that the prototype is capable of generating is limited to a pre-defined range of feedback types, including hints, explanations, examples, answers, and code snippets. In this respect, the prototype's capability to select the type of feedback is based on a pre-specified range of possible instructional actions that the prototype is capable of generating.